\documentclass[iop]{emulateapj} 
\usepackage{color} 
\usepackage{epsfig}
\usepackage{graphicx}
\usepackage[colorlinks=true, linkcolor=blue, citecolor=blue, urlcolor=magenta]{hyperref}

\shorttitle{Target Selection for the LBTI-HOSTS Survey} 
\shortauthors{Weinberger et al.}
 
\begin{document} 

\journalinfo{Accepted to ApJ Suppl. Series., December 12, 2014}
\submitted{}
 
\title{Target Selection for the LBTI Exozodi Key Science Program} 
 
\author{Alycia J. Weinberger}
\affil{Department of Terrestrial Magnetism, Carnegie Institution for
  Science, 5241 Broad Branch Road NW, Washington, DC 20015, USA, weinberger@dtm.cwi.edu}
\author{Geoff Bryden}
\affil{Jet Propulsion Laboratory, California Institute of
  Technology, 4800 Oak Grove Dr, Pasadena, CA 91109, USA}
\author{Grant M. Kennedy}
\affil{Institute of Astronomy, University of Cambridge, Madingley
  Road, Cambridge CB3 0HA, UK}
\author{Aki Roberge}
\affil{Exoplanets \& Stellar Astrophysics Laboratory, NASA Goddard
  Space Flight Center, Code 667, Greenbelt, MD 20771, USA}
\author{Denis Defr\`ere, Philip M. Hinz}
\affil{Steward Observatory, University of Arizona, 933 N. Cherry Lane,
  Tucson, AZ, 85721, USA}
\author{Rafael Millan-Gabet}
\affil{NASA Exoplanet Science Institute, California Institute of Technology,
  Pasadena, CA, 91125, USA}
\author{George Rieke, Vanessa P. Bailey}
\affil{Steward Observatory, University of Arizona, 933 N. Cherry Lane,
  Tucson, AZ, 85721, USA}
\author{William C. Danchi}
\affil{Exoplanets \& Stellar Astrophysics Laboratory, NASA Goddard
  Space Flight Center, Code 667, Greenbelt, MD 20771, USA}
\author{Chris Haniff}
\affil{Cavendish Laboratory, University of Cambridge, JJ Thomson Avenue, Cambridge CB3 0HE, UK}
\author{Bertrand Mennesson, Eugene Serabyn}
\affil{Jet Propulsion Laboratory, California Institute of
  Technology, 4800 Oak Grove Dr, Pasadena, CA 91109, USA}
\author{Andrew J. Skemer}
\affil{Steward Observatory, University of Arizona, 933 N. Cherry Lane,
  Tucson, AZ, 85721, USA}
\author{Karl R. Stapelfeldt}
\affil{Exoplanets \& Stellar Astrophysics Laboratory, NASA Goddard
  Space Flight Center, Code 667, Greenbelt, MD 20771, USA}
\author{Mark C. Wyatt}
\affil{Institute of Astronomy, University of Cambridge, Madingley
  Road, Cambridge CB3 0HA, UK}

\begin{abstract} 

The Hunt for Observable Signatures of Terrestrial planetary Systems (HOSTS) on
the Large Binocular Telescope Interferometer will survey nearby stars for
faint emission arising from $\sim$300~K dust (exozodiacal dust), and aims to
determine the exozodiacal dust luminosity function. HOSTS results will enable
planning for future space telescopes aimed at direct spectroscopy of habitable
zone terrestrial planets, as well as greater understanding of the evolution of
exozodiacal disks and planetary systems.  We lay out here the considerations
that lead to the final HOSTS target list. Our target selection strategy
maximizes the ability of the survey to constrain the exozodi luminosity
function by selecting a combination of stars selected for suitability as
targets of future missions and as sensitive exozodi probes.  With a survey of
approximately 50 stars, we show that HOSTS can enable an understanding of the
statistical distribution of warm dust around various types of stars and is
robust to the effects of varying levels of survey sensitivity induced by
weather conditions.
\end{abstract} 
 
 
 
\section{Introduction} \label{sec:intro}
The Hunt for Observable Signatures of Terrestrial planetary Systems (HOSTS) on
the Large Binocular Telescope Interferometer (LBTI) will survey nearby stars
for faint exozodiacal dust (exozodi). This warm circumstellar dust, such as
that found in the vicinity of Earth, is generated in asteroidal collisions and
cometary breakups. We define exozodiacal dust as sitting in the habitable
zone, that is $\sim$1 AU from a Solar-type star, and therefore as having a
temperature comparable to the Earth, i.e. $\sim$278 K.

The goal of the LBTI HOSTS survey is to provide information on exozodi needed
to develop a future space telescope aimed at direct detection of habitable
zone terrestrial planets (aka. exoEarths).  The habitable zone is defined by
where a terrestrial planet can have long-term surface water, but its exact
boundaries depend on planetary properties.  Nevertheless, surface temperatures
near 300~K imply that Earth-mass exoplanets need insolations comparable to
that of Earth up to 1.2 times greater than Earth's
\citep[e.g.][]{Leconte:2013,Kopparapu:2013}.  There is no single agreed upon
definition of exozodi in the literature \citep{Roberge:2012}. The HOSTS team
has adopted a definition that scales the surface density of the Sun's Zodiacal
disk at the Earth equivalent insolation distance (EEID). Thus the surface
density profile expands with stellar luminosity, and allows the ``exozodi''
level to be compared across stars of different types. See the companion paper
\citet{Kennedy:2014} for a full discussion of our adopted model. This
reference model includes dust interior to the habitable zone all the way in to
the sublimation radius, so this model may test how close-in dust such as that
detected in near-infrared interferometric surveys
\citep{Absil:2013,Ertel:2014} is related to habitable zone dust.

The typical exozodi detection from space-based photometry and
spectrophotometry, primarily with the IRS instrument on the Spitzer Space
Telescope, is $\sim$1000 times the Solar System's level (3 $\sigma$),
i.e. 1000 zodi \citep{Beichman:2006,Lawler:2009,Chen:2014}.  The best limits
from the ground-based Keck interferometer are 500 zodi (3$\sigma$)
\citep{Millan-Gabet:2011,Mennesson:2014}. Interferometric searches for dust in
the near-infrared can find dust interior to the habitable zone, at temperatures
$\gtrsim$500K \citep{Absil:2013,Ertel:2014} or that comes from scattered light,
and far-infrared and submillimeter telescopes can find dust much cooler than
exozodi, at temperatures $<$100~K \citep[e.g.][]{Eiroa:2013}.  LBTI-HOSTS will
be the first survey capable of measuring exozodi known to be at habitable zone
temperatures and at the 10-20 zodi level (3$\sigma$).

Exozodi of this brightness would be the major source of astrophysical noise
for a future space telescope aimed at direct imaging and spectroscopy of
habitable zone terrestrial planets.  For example, more than about 4 zodis
would cause the integration time for coronagraphic imaging of an Earth-like
planet in the habitable zone of a G2V star at 10 pc to exceed 1 day, using a
4m telescope and the other baseline astrophysical and mission parameters given
in \citet{Stark:2014}. A larger telescope can tolerate larger zodi levels for
the same integration time.

Detections of warm dust will also reveal new information about planetary
system architectures and evolution. Asteroid belts undergoing steady state
collisions should grind themselves down in much less time than the Gyr ages of
nearby stars.  So, warm debris disks around old stars may signal late cometary
influxes or stochastic collisional events
\citep[e.g.][]{Wyatt:2007,Gaspar:2013}.  While $\sim$20\% of nearby stars
have cold, i.e. $<$150~K, dust \citep{Eiroa:2013} and $\sim$15\% have hot,
i.e. $>$500~K, dust \citep{Ertel:2014}, there is presently no demonstrated connection
between the two. To understand the evolution of planetary
systems, we seek to measure the luminosity function of exozodi with age and
stellar mass and determine whether the presence of cold outer disks correlates
with warm inner exozodi.

LBTI is a nulling interferometer, designed to use the 8.4 m apertures of the
LBT fixed in a common mount at a 14.4 m separation, for the detection of
emission from warm dust around nearby stars.  LBTI works in the thermal
infrared, employing dual adaptive secondaries to correct atmospheric seeing,
and providing low thermal background and high Strehl images to the science
camera NOMIC \citep{Hinz:2008,Hinz:2012}.  Closed loop phase tracking in the
near infrared is used to stabilize the destructive interference of the star at
N-band (9.8-12.4 $\mu$m) and detect flux from the resolved dust disk \cite{Defrere:2014a}.  The
separation of the LBT mirrors at a working wavelength of 11 $\mu$m produces a
first transmission peak centered at 79 mas (1 AU at 13 pc) and an inner
working angle (half transmission) of 39 mas (1 AU at 25 pc).

Together, observations of thermal emission from disks with LBTI and
images with space-based optical coronagraphs capable of probing the same
angular scales in scattered light will measure the albedo of dust
grains. Albedo is one of the few available constraints on dust composition and
thereby parent body composition for debris disks. Scattered light images of
dust in the habitable zones of several nearby stars may be possible with a
coronagraph on the WFIRST-AFTA mission \citep{Spergel:2013}.

\section{Target List Assembly and Exclusions}

\subsection{Target Selection Goals}

Target selection for HOSTS is a balance between including stars that are expected targets
of a future exoEarth mission and including stars of various types to enable the best
understanding of the statistical distribution of exozodi over a range of parameters.  The
two approaches are complementary and together enable investigations of habitable zone dust
production across a range of host stellar types.

The mission-driven approach concentrates on F, G, and K-type stars that are
the best targets for future direct observations of exoEarths, thereby
providing ``ground truth'' dust observations. The sensitivity sweet spot for
an optical planet imager lies with G and K stars because 1) the planet-to-star
contrast ratio is inversely proportional to stellar luminosity and 2) the
orbital radius of the habitable zone increases as $\sqrt{L_*}$
\footnotemark[1]. As a result, M-type stars have favorable planet-to-star contrast
ratios but habitable zones close to the stars, whereas A-type stars have poor
contrast ratios and habitable zones further from the stars.

\footnotetext[1]{The EEID, i.e. where a planet receives the same incident flux
  as Earth, defines the habitable zone (see Section \ref{sec:sunlike}). Since
  the flux at the EEID is a constant, a 1~R$_\earth$ planet there always has
  the same absolute magnitude independent of host star luminosity.  However,
  the absolute magnitude of stars decreases toward earlier spectral type
  stars, thus increasing the star-to-planet flux ratio.  The radial
  temperature dependence of a blackbody emitter in a stellar radiation field
  can be calculated by equating the incident flux (L$_*$/4$\pi$) with the
  emergent flux (4$\sigma$r$_{\rm EEID}^2$T$_{\rm HZ}^4$). Thus, for a
  fixed temperature, as in a habitable zone, the radius at which a blackbody
  reaches that temperature is proportional to $\sqrt{L}$.}

Not every potential target of a future exoEarth mission can be observed with
LBTI; for one thing, many lie in the southern hemisphere and are not
observable from LBT on Mount Graham, AZ.  Furthermore, some stars bright
enough at visual wavelengths and therefore accessible to an exoEarth mission
would be too faint for LBTI to achieve good sensitivity in the limited total
observing time.  Our goal is to design a survey that can fully inform target
selection for a future exoEarth mission; survey results will have to be
modeled and then extrapolated to lower dust levels.  Therefore, there must be
observational extensions to the mission-driven sample that will inform models
of dust evolution and aid extrapolation.

The second approach, a LBTI sensitivity-driven approach, selects targets based
only on the expected LBTI exozodi sensitivities, without consideration of
exoEarth mission constraints.  This would naturally select more early-type
stars (A stars and early F-type stars) because they are brighter, have
habitable zones at large separations, and higher F$_{\rm disk}$/F$_*$ at
N-band (see \citet{Kennedy:2014} for details). Therefore, the results of this
type of survey would have to be extrapolated to later spectral type targets
using planet formation theory.

The brightest nearby late-F to K-type stars can satisfy both the mission and
sensitivity-driven selection criteria, and we give a description of these in Section
\ref{sec:sunlike}, we show that there are 25-48 such stars, depending on LBTI sensitivity.
We anticipate that HOSTS will survey $\sim 50$ stars, given the amount of observing time
allocated on LBTI, so the target selection approach followed will determine the rest of
the observed stars.

We lay out here the considerations that lead to the final HOSTS target
list. We discuss how to balance mission-driven and sensitivity considerations
to maximize scientific return from the HOSTS project. By presenting our target
list in this early paper, we also hope to encourage intensive study of these stars
with other techniques that will eventually enhance our ability to understand the
evolution of circumstellar dust with time.

\subsection{Target Selection Constraints} \label{sec:constraints}

We started with a list of all bright, northern main sequence stars of spectral
types A through M observable from LBT (declination $> -30^\circ$) by using two
catalogs: the Unbiased Nearby Stars (UNS) sample assembled for cold debris
disks studies \citep{Phillips:2010} and the Hipparcos 30 pc sample assembled
for exoEarth mission planning \citep{Turnbull:2012}. UNS is complete to about
16 pc for K-type stars and about 45 pc for A-type.

Binary stars were excluded based on both technical and scientific
criteria. There are two technical reasons to exclude binary stars: 1) to
ensure that the adaptive optics system can easily lock on the target of
interest and 2) to ensure that flux from the companion does not contaminate
the area being searched for exozodi emission.  We therefore excluded binary
stars with separations $< 1\farcs5$. Some stars are known to be spectroscopic
binaries (SBs) but without well-measured orbits. We excluded all such SBs
because their maximum separations might fall within an angular range of 10s to
100s of mas and provide confusing non-null signals. The main sources of
information about multiplicity were the Washington Visual Double Star Catalog
\citep{Mason:2013} and the 9th Catalogue of Spectroscopic Binary Orbits
\citep{Pourbaix:2009}.

We further excluded stars with flux densities $<$1 Jy in the broad N-band
($\sim$11 $\mu$m) filter used for the HOSTS survey.  We anticipate that the
LBTI null would be degraded for fainter stars. To estimate the brightness of
our targets, we fit Kurucz stellar models to available photometry at BVJHK
plus WISE bands W3 (12 $\mu$m) and W4 (22 $\mu$m) and then used the model to
predict the NOMIC flux density.

We also only considered stars with inner habitable zone distances probed by
the LBTI transmission pattern, i.e. zones larger than about 60 mas. An exozodi
disk smaller than this has low transmission efficiency, i.e. it is nulled
along with the star because the LBTI transmission pattern peak is at $\approx$79 mas.
A general result of our brightness cuts is that our target
stars are all within 28 pc. Therefore, our angular criterion, above, excluded
binaries with separations $\lesssim$50 AU. Furthermore, studies of
protoplanetary disk evolution indicate that stellar companions within 100~AU
of the primary stars cause lower disk masses and faster disk dissipation,
possibly inhibiting planet formation
\citep[e.g.][]{Osterloh:1995,Jensen:1996,Andrews:2005,Harris:2012}. We therefore also excluded physical
binaries with separations $<$100~AU. Although it would be interesting to study
the effect of binary separation on habitable zone dust, we emphasized the
formation of an overall sample with as few selection effects as possible and
eschewed inclusion of subsamples too small to provide statistically meaningful
results.

Finally, we excluded giant stars (luminosity class III), i.e. stars that
appear significantly brighter than the main sequence. LBTI would probe regions
around these stars that are significantly larger than the size of the
habitable zones that existed when the stars resided on the main sequence and
thus not directly comparable to the rest of the sample.  Table
\ref{tab:binary} lists the targets excluded for binarity and location above
the main sequence.

\subsection{Target List Categories}

We categorize the targets that meet the above criteria into two samples
described below in:

Section \ref{sec:sunlike}: The Sun-like sample includes targets with spectral types
later than F5.  These 48 stars are potential targets for a future exoEarth
mission. Of these 25 have flux density $>$2 Jy at N-band.

Section \ref{sec:bright}: The sensitivity-driven, i.e. early-type star, sample
includes targets with spectral types between A0 and F4.  These 20 stars provide
additional information on the exozodi luminosity function. Of these, 15 have
flux density $>$2Jy at N-band.
 
Together, there are 68 sources in the above categories from which the
optimal HOSTS survey can be created.


\section{Sun-like Sample} \label{sec:sunlike} 

Our objective for this LBTI-HOSTS sub-sample is to observe stars that are probable targets
for a future exoEarth mission, based on current knowledge, and stars that inform our
understanding of the typical exozodi levels around similar stars.  These observations will
provide dust measurements (or upper limits) for specific stars.  They will also supply a
larger sample of solar-type stars with which to model the distribution of exozodi levels
for Sun-like stars. This will enable evaluation of the suitability, as exoEarth mission
targets, of individual stars that could not be observed from LBTI (because, for example,
they were too faint or too far south). Here, we define ``Sun-like'' as having spectral
types later than or equal to F5. The coolest star that passed all our cuts is spectral
type K8. The majority of high photometric quality targets for the Kepler mission's
exoplanet search are also of spectral types mid-K to mid-F (4500-7000 K)
\citep{Christiansen:2012}.

The great technical challenge for direct exoEarth observations is to suppress
the central starlight tremendously yet allow light from extremely faint
planets to be detected at small angular separations.  Therefore, the best
systems to search for exoEarths are those with widely separated habitable
zones (HZs) and with high planet-to-star flux ratios.  A full discussion of
all the considerations that go into determining a star's habitable zone
boundaries appears in \citet{Kopparapu:2013}.  However, to first order, the
location of a star's HZ is set by how much light would hit an Earth-twin
planet. Therefore, the Earth-equivalent insolation distance (EEID)
approximately scales in the following way
\begin{equation}
r_\mathrm{EEID} \: \approx \: r_\oplus \times \left( L_\star \, / \, L_\sun \right)^{1/2 } \; ,
\end{equation}
where $L$ is the bolometric luminosity and $r_\oplus$ is the Earth-Sun distance.

Following \citet{Turnbull:2012}, the planet-to-star reflected flux ratio at
visible wavelengths for
an Earth-twin planet is approximately
\begin{equation}
\left(F_p \, / \, F_\star \right)_\mathrm{HZ} \: \approx \: (1.2 \times 10^{-10}) \, / \, (L_\star/L_\odot)
\end{equation}
So as the stellar luminosity increases, the HZ moves
outwards, increasing the separation of an exoEarth from its host star (good).
However, simultaneously the planet-to-star flux ratio decreases, resulting in
longer exposure times to reach a given detection limit (bad).

These two competing effects largely dictate which stars are the best for
direct observations of exoEarths.  The current consensus is that starlight
suppression technologies working at optical wavelengths (e.g.\ internal
coronagraphs) are the most advanced \citep{Greene:2013}. For these mission
concepts, the best targets are nearby stars of mid-F, G, and K spectral types.
In general, for a given optical coronagraphic telescope aperture, the less
exozodi noise contribution that a star system has, the earlier the spectral type
of star systems that can be searched with high completeness.  An interferometric
mission, such as an array of 4$\times$4m free-flying mid-infrared telescopes,
provides somewhat different completeness as a function of stellar
luminosity. For the HOSTS survey, we make no assumptions about exoEarth
detection technology.  If we keep the ratio of F:G:K stars fixed, the best
targets for an interferometric telescope agree well with the coronagraphic
telescope list \citep{Defrere:2010}.

\begin{figure}
\centering\epsfig{file=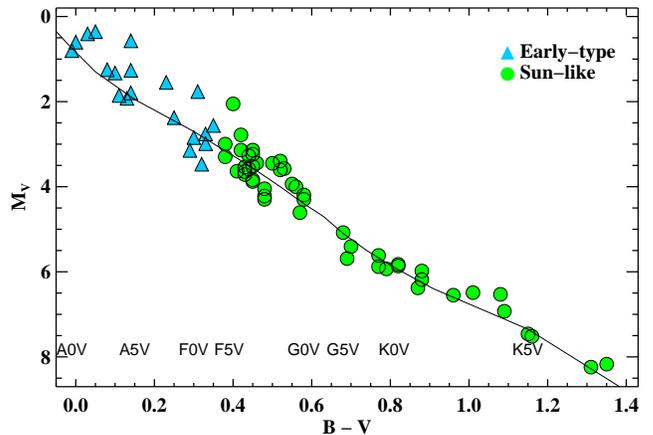,width=2.25in,clip=,angle=90}
\caption{Color magnitude (absolute V magnitude versus B-V color) plot of the
  complete sample. Sun-like stars, defined as spectral type F5 and later, are
  shown with green filled circles. Early type stars, defined as spectral types F4 and
  earlier, are show with blue filled triangles. The black line shows the MK main sequence
  as given in \citet{DrillingLandolt}. \label{fig:cmd}}
\end{figure}

\begin{figure}[htb]
\centering\epsfig{file=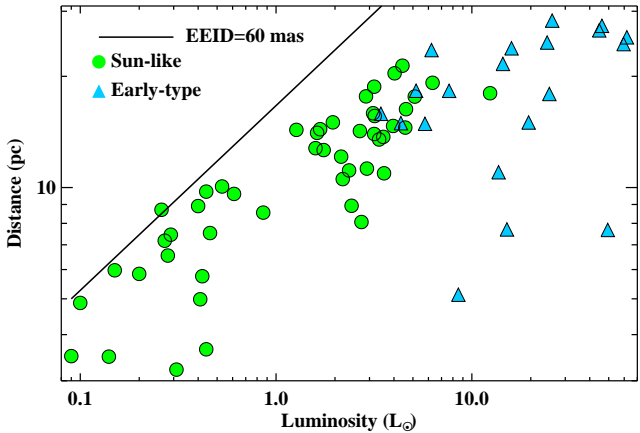,width=2.25in,clip=,angle=90}
\caption{Distance versus luminosity for the complete sample. The black line shows an Earth
  equivalent insolation distance (EEID) of 60 mas.  Stars that fall above this line were
  excluded because their exozodi would largely fit within the first null, and therefore
  LBTI would not be very sensitive to such dust. The LBTI the inner working angle in
  N-band (11 $\mu$m), defined as $\lambda$/4B, is $\approx$ 39 mas for the LBT mirror
  separation of 14.4 m while the first transmission peak is at 79 mas.  That stars $>$1
  L$_\odot$ fall well below the line shows that the N-band flux density requirement drives
  the source selection rather than the EEID requirement. \label{fig:eeid}}
\end{figure}

We found 48 stars that met all of our selection criteria and some of their
basic properties are listed
in Table \ref{tab:sunlike} and shown in Figures \ref{fig:cmd} and
\ref{fig:eeid}. Our current working knowledge of the LBTI system is that the
null quality will not depend on stellar brightness for stars brighter than 2
Jy at N-band; there are 25 such bright stars on our list. We expect that for
stars fainter than 2 Jy, the degradation in the null will be a gentle function
of brightness, but this remains to be tested.

The mean distance to these sample stars is 11.4~pc; the closest star is at
3.2~pc and the most distant at 21.4~pc.  The presence/absence of a disk was
not a criterion for selecting stars in the Sun-like star sample. What is known
about the presence or absence of hot/warm (potentially N-band detectable) and cold (far-IR
detectable) circumstellar disks is noted in the Table. Each dust survey has
somewhat different limits; the reader should consult the original papers for
details.

\section{Sensitivity-Driven, i.e. Early-type, Sample} \label{sec:bright} 

Our objective for this sample is to find stars for which LBTI can make its most sensitive
observations of $\sim$300~K dust, regardless of the spectral type of the host
star.  To create this sample, we select all stars that have N-band flux densities $\geq
1$~Jy and for which the location of the EEID is $>$60 mas. This preferentially selects
A-type and early F-type stars.  In general, these are not good exoEarth imaging targets
themselves, because of the low habitable-zone-planet-to-star contrast. However, they will
provide an important addition to our understanding of the exozodiacal dust luminosity
function as it might depend on mass and luminosity of the host star.

We find an additional 20 stars that meet our selection criteria and were not
already selected in the Sun-like Sample. These stars are all earlier spectral
type than F4 and are given in Table~\ref{tab:bright}.  These stars are
typically further away than the Sun-like samples tars, with an average
distance of 18.6~pc.  Twelve stars have significant infrared excesses
indicating abundant circumstellar dust at some distance from the stars;
references are given in the table.


\section{Discussion} \label{sec:discussion} 

Despite our attempts to reject known binaries (see Section
\ref{sec:constraints}), there could be unknown companions that would transmit
through the LBTI null pattern and therefore generate some signal. There are
some ways to distinguish a companion from a disk using LBTI. A companion will
transmit primarily through a single LBTI fringe, unlike a spatially resolved
disk. Therefore, a companion will produce a null that varies as the
orientation of the projected baseline of the interferometer changes due to
Earth's rotation over an observation. However, an inclined disk would have a
similar effect; therefore distinguishing a companion from a disk will likely
require follow-up observations. For example, measuring the source null in
narrower filters at the short and long wavelength ends of N-band, i.e. 8 and
12.5 $\mu$m, would provide some information on its temperature and spatial
extent.  Radial velocity observations will constrain the possible masses and
orbital periods of suspected companions.

Any companions discovered by LBTI are likely to be of substellar mass.  All
but four of the Sun-like sample stars and seven of the Early-type sample stars
have been studied extensively by radial velocity planet search programs
\citep[e.g.][]{Butler:2006,Lagrange:2009,Fischer:2014}. At the separation of
maximum transmission, i.e. 79 mas, a 80 M$_{\rm Jup}$ brown dwarf in an orbit
inclined at 45$^\circ$ would induce a typical reflex motion of about 2~km
s$^{-1}$ for our sample stars, which could be detected for all but the most
rapidly rotating stars in the sample.

\begin{figure}[h]
\centering\epsfig{file=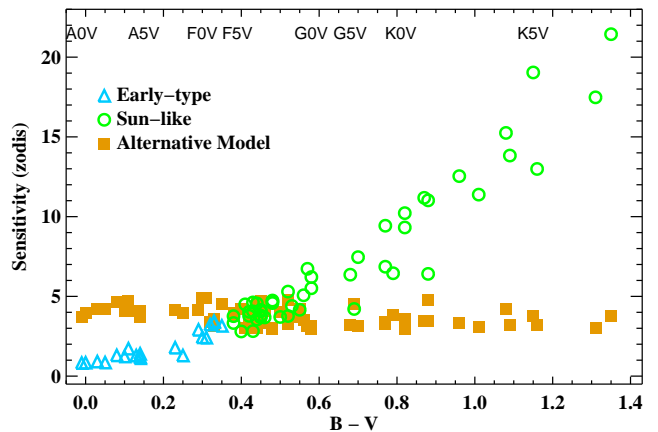,width=2.25in,clip=,angle=90}
\caption{Expected LBTI sensitivity (1 $\sigma$) to dust, in zodis as defined in
  \citet{Kennedy:2014} and assuming a null depth of 10$^{-4}$ for the complete sample
  (green and blue circles). In this model, the surface density of the disk in the
  habitable zone is fixed and the flux density of the disk relative to the
  star is $\propto T^3_\star$, so LBTI is more sensitive to dust around hotter (bluer) stars.
Note that these values of the sensitivity assume that the null is limited by
  systematics and not by target flux density. An alternative definition of a
  zodi is a fixed flux density in the habitable zone (orange squares); in this case
  LBTI's limits are only weakly a function of stellar luminosity.  \label{fig:zodisens}}
\end{figure}

In advance of scheduled HOSTS observing runs, a prioritized list of targets will be
constructed based on the target observability (e.g. above airmass 1.5 for more than 2 hr)
and our expected sensitivity to exozodi.  To determine our expected sensitivity, we pass
an exozodi model, described in \citet{Kennedy:2014}, through an LBTI null model to
calculate an exozodi limit for each star, in the unit of a ``zodi.'' This model was
designed to be simple, to facilitate comparisons between observations of stars of varying
properties, and to have a straightforward correspondence with the Solar System's actual
Zodiacal cloud. The basic features of this model are a fixed surface density in the
habitable zone, i.e. at a temperature of 278~K, and a weak power law dependence of surface
density on radius from the star that matches the Zodiacal cloud's.  
Figure \ref{fig:zodisens} shows this estimation based on current knowledge
of the achievable LBTI null depth. LBTI is still in commissioning, so the final dependence
of null depth on target brightness is not yet well established. We have assumed that for
targets brighter than 2 Jy, LBTI will be systematics limited, so the target's flux density
will not affect our zodi sensitivity. However, there may be additional degradation of the
null for targets of 1--2 Jy, which comprise 28/68 of our targets.

There are many other possible definitions of a ``zodi'' including ones defined in terms of
a fixed $F_{\rm dust}/F_\star$ or $L_{\rm dust}/L_\star$ at a given temperature
\citep{Roberge:2012}.  For comparison purposes, we also calculate an exozodi sensitivity
for each of our target stars by assuming a version of our reference model that contains
dust only in the habitable zone, i.e. extending from 0.95 - 1.37 $\sqrt{L_*}$ and
normalized to a fixed $F_{\rm dust}/F_\star =  5 \times 10^{\rm -5}$ at a wavelength of
11 $\mu$m (for our NOMIC filter). These limits are also shown in Figure
\ref{fig:zodisens}. 

The goal of the overall survey is not only to identify exozodiacal dust around specific
stars of interest, but also to measure the luminosity function of disks in a general
statistical sense.  As such, we define here a key metric for the overall survey - $Z10$ -
the fraction of stars with more than 10 zodis. This level of exozodi, versus a Solar
System level of dust, would cut the number of Earth-like planets imaged by a future
direct-imaging mission by half \citep{Stark:2014}.  This recent work also shows, however,
that a mission that observes an ensemble of stars has a total planet yield that is a weak
function of the exozodi level \citep{Stark:2014}.

In a real-world ground-based observing program, under changing seeing and
transparency and seasonally biased conditions, it will be impossible to
observe all stars, even those brighter than 2 Jy, to identical zodi detection
depths.  Of the 68 stars in Tables 1 and 2, we expect to observe $\sim$50.
What is critical is that no biases to the sample are introduced during the
selection of the actual observed targets.

The ability of the LBTI survey to constrain $Z10$ depends on both the number of observed
targets and the sensitivity of each individual measurement.  We performed Monte Carlo
simulations to estimate the expected accuracy of $Z10$ as a function of the number of
targets.  Assuming that the underlying distribution of disk brightnesses follows a
log-normal distribution whose width is set by $Z10$, we determine how well $Z10$ is
constrained by the LBTI observations.  We assume that each star is treated as a unique
observation. The bright end of the distribution is already constrained by Spitzer/KIN/WISE
observations \citep{Lawler:2009, Millan-Gabet:2011, Kennedy:2013}; therefore, we set the
frequency of 1000 zodi disks to be 1\%.

\begin{figure}[th]
\begin{center}   	
\includegraphics[width=3.in]{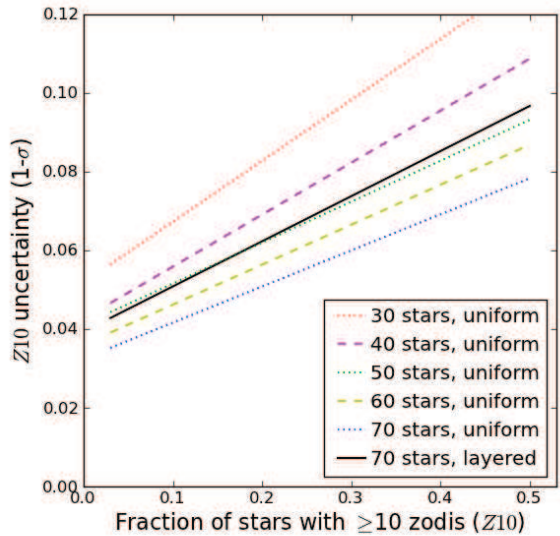}
\end{center}
\caption{The ability of the overall survey to constrain $Z10$ (the fraction of
  stars with $\geq$ 10 zodis of dust) depends on the number of targets and the
  accuracy of each measurement.  Based on Monte Carlo simulations we find that
  a survey with two levels of sensitivity - 40 stars with 3-zodi accuracy and
  30 stars with 10-zodi accuracy (black line) - is roughly equivalent to a 50
  star survey of uniform 3-zodi depth (green dashed line).
}\label{layeredFig}
\end{figure}

At first we consider a uniform survey depth with 3-zodi sensitivity for each
measurement (1-$\sigma$), which we assume would be the perfect, likely
unachievable, survey LBTI could perform.  Figure~\ref{layeredFig} shows how
well $Z10$ is constrained by uniform-depth surveys ranging from 30 to 70
stars.  We find that a 50 star survey can measure $Z10$ with $\sim$20\%
accuracy (for $Z10$$\simeq$0.3-0.4).  Using advanced statistical methods to
allow averaging over multiple targets to achieve deeper zodi limits, it may be
possible to improve on these rough limits \citep{Mennesson:2014}.

Since variations in weather will inevitably result in non-uniform sensitivity,
Figure~\ref{layeredFig} also shows the constraints on $Z10$ for a 2-layered
survey, where 40 stars are observed with 3-zodi accuracy and another 30 stars
with only 10-zodi accuracy.  We find that this layered survey has equivalent
power to reveal the zodi luminosity function as a 50 star survey done to a
depth of 3 zodis.  We conclude that an optimal observing strategy should not
mandate uniform sensitivity, thereby concentrating a large fraction of
telescope time on a small number of stars, but will instead observe a greater
number of stars, some with greater depth than others.

The HOSTS survey is expected to begin in 2015 and to continue for two to three years. 
During commissioning, LBTI observed $\eta$ Crv, one of the early-type sample
stars with a known mid-infrared excess. The observations demonstrate the power
of LBTI to constrain the dust distribution in the habitable zone
\citep{Defrere:2014}.

\acknowledgements 

The Large Binocular Telescope Interferometer is funded by the National Aeronautics and
Space Administration as part of its Exoplanet Exploration Program.  This work of GMK \&
MCW was supported by the European Union through ERC grant number 279973. This research has
made use of the SIMBAD database and the VizieR catalogue access tool, CDS, Strasbourg,
France and the Washington Double Star Catalog maintained at the U. S. Naval Observatory.
  
\section{Appendix: Binary Stars Excluded from Sample \label{tab:binaries}}

We list here (Table 3) stars that could otherwise meet our selection criteria but that
were excluded due to binarity as described in Section 2.2. Much of
the information in this table comes from Washington Double Star Catalog (Mason
et al. 2001-2014).


\begin{deluxetable*}{llrrccccll}
\tabletypesize{\footnotesize}
\tablecaption{Stars in the Sun-like Star Sample \label{tab:sunlike}} 
\tablewidth{0pt}
\tablehead{
\colhead{HD}&\colhead{Name}&\colhead{RA} &\colhead{Dec}&\colhead{Distance}
                                                   &\colhead{Spectral} &\colhead{EEID}
                                                          &\colhead{F$_\nu$ (N-band)}  &\colhead{hot/warm}&\colhead{cold}\\
            &              &              &    &\colhead{(pc)}&\colhead{Type}&  &\colhead{(Jy)}&\colhead{excess}  &\colhead{excess}
}
\startdata
693     &6 Cet          & 00:11:15.86  &-15:28:04.7  &18.7  & F8V     &0.095  &1.15  &        &n (E13)\\
4628    &               & 00:48:22.98  &+05:16:50.2  &7.5   & K2.5V   &0.072  &1.30  &        &n (T08)\\
9826    &$\upsilon$ And & 01:36:47.84  &+41:24:19.6  &3.5   & F9V     &0.136  &2.36  &n (A13) &n (B06,E13)\\ 
10476   &107 Psc        & 01:42:29.76  &+20:16:06.6  &7.5   & K1V     &0.090  &2.02  &n (MG11)&n (T08)\\     
10700   &tau Cet        & 01:44:04.08  &-15:56:14.9  &3.6   & G8.5V   &0.182  &5.42  &y (A13) &y (G04)\\     
10780	&GJ 75	        & 01:47:44.83  &+63:51:09.0  &10.1  & K0V     &0.072  &1.12  &        &n (L09)\\
16160   &GJ 105         & 02:36:04.89  &+06:53:12.7  &7.2   & K3V     &0.073  &1.53  &        &n (T08)\\
16895   &13 Per         & 02:44:11.99  &+49:13:42.4  &11.1  & F7V     &0.138  &2.43  &n (A13) &n (Be06)\\    
17206   &tau01 Eri      & 02:45:06.19  &-18:34:21.2  &14.2  & F75     &0.115  &1.69  &        &n (T08)\\
19373   &iot Per        & 03:09:04.02  &+49:36:47.8  &10.5  & F9.5V   &0.141  &2.85  &n (MG11)&n (T08) \\ 
22049   &eps Eri        & 03:32:55.84  &-09:27:29.7  &3.2   & K2V     &0.172  &7.39  &n (A13) &y (B09)\\ 
22484   &LHS 1569       & 03:36:52.38  &+00:24:06.0  &14.0  & F8V     &0.127  &2.35  &y (A13) &y (T08)\\
23754   &tau06 Eri      & 03:46:50.89  &-23:14:59.0  &17.6  & F5IV-V  &0.128  &2.10  &        &n (G13)\\
26965   &omi Eri        & 04:15:16.32  &-07:39:10.3  &5.0   & K0.5V   &0.128  &3.51  &        &n (L02)\\ 
30652   &1 Ori          & 04:49:50.41  &+06:57:40.6  &8.1   & F6V     &0.205  &4.76  &n (A13) &n (T08)\\          
32147   &               & 05:00:49.00  &-05:45:13.2  &8.7   & K3V     &0.059  &1.00  &        &n (L09)\\
34411   &lam Aur        & 05:19:08.47  &+40:05:56.6  &12.6  & G1.5IV  &0.105  &1.80  &n (MG11)&n (T08)\\         
35296   &V1119 Tau      & 05:24:25.46  &+17:23:00.7  &14.4  & F8V     &0.090  &1.03  &        &n (T08)\\
38393   &gam Lep        & 05:44:27.79  &-22:26:54.2  &8.9   & F6V     &0.175  &4.40  &n (MG11)&n (Be06)\\  
48737   &ksi Gem        & 06:45:17.36  &12:53:44.13  &18.0  & F5IV    &0.196  &4.34  &n (A13) &n (K13) \\
78154   &sig02 Uma A    & 09:10:23.54  &+67:08:02.4  &20.4  & F6IV    &0.099  &1.24  &        &n (G13)\\
84117   &GJ 364         & 09:42:14.42  &-23:54:56.0  &15.0  & F8V     &0.093  &1.11  &        &n (E13)\\
88230   &NSV 4765       & 10:11:22.14  &+49:27:15.3  &4.9   & K8V     &0.065  &1.91  &n (MG11)&n (T08)\\         
89449   &40 Leo         & 10:19:44.17  &+19:28:15.3  &21.4  & F6IV    &0.098  &1.10  &        &n (G13)\\
90839   &36 Uma         & 10:30:37.58  &+55:58:49.9  &12.8  & F8V     &0.099  &1.25  &        &n (T08)\\
95128   &47 Uma         & 10:59:27.97  &+40:25:48.9  &14.1  & G1V     &0.091  &1.35  &n (MG11)&n (T08)\\
101501  &61 Uma         & 11:41:03.02  &+34:12:05.9  &9.6   & G8V     &0.081  &1.24  &        &n (G03)\\
102870  &bet Vir        & 11:50:41.72  &+01:45:53.0  &10.9  & F9V     &0.173  &4.30  &n (A13) &n (T08)\\
115617  &61 Vir         & 13:18:24.31  &-18:18:40.3  &8.6   & G7V     &0.108  &2.20  &n (L09,E14) &y (L09)\\ 
120136  &tau Boo        & 13:47:15.74  &+17:27:24.8  &15.6  & F6IV    &0.114  &1.67  &n (E14) &n (B09)\\
126660  &tet Boo        & 14:25:11.8   &+51:51:02.7  &14.5  & F7V     &0.147  &3.12  &        &n (T08)\\
131977  &KX Lib         & 14:57:28.00  &-21:24:55.7  &5.8   & K4V     &0.076  &1.95  &n (L09) &n (Be06) \\ 
141004  &lam Ser        & 15:46:26.61  &+07:21:11.0  &12.1  & G0IV-V  &0.121  &2.40  &n (A13) &n (K10)\\ 
142373  &LHS  3127      & 15:52:40.54  &+42:27:05.5  &15.9  & F8Ve    &0.111  &2.03  &n (A13) &n (T08)\\
142860  &gam Ser        & 15:56:27.18  &+15:39:41.8  &11.2  & F6IV    &0.151  &2.93  &n (A13) &n (T08)\\
149661  &V2133 Oph      & 16:36:21.45  &-02:19:28.5  &9.8   & K2V     &0.068  &1.00  &n (E14) &n (T08)\\
156026  &V2215 Oph      & 17:16:13.36  &-26:32:46.1  &6.0   & K5V     &0.064  &1.64  &        &n (Be06)\\
157214  &w Her          & 17:20:39.30  &+32:28:21.2  &14.3  & G0V     &0.079  &1.01  &        &n (T08)\\
160915  &58 Oph         & 17:43:25.79  &-21:40:59.5  &17.6  & F5V     &0.096  &1.18  &n (E14) &n (E14)\\
173667  &110 Her        & 18:45:39.72  &+20:32:46.7  &19.2  & F6V     &0.131  &2.18  &y (A13) &n (T08)\\
185144  &sig Dra        & 19:32:21.59  &+69:39:40.2  &5.8   & G9V     &0.113  &2.72  &n (A13) &n (T08)\\
192310  &GJ 785         & 20:15:17.39  &-27:01:58.7  &8.9   & K2+V    &0.071  &1.25  &        &n (Be06)\\
197692  &psi Cap        & 20:46:05.73  &-25:16:15.2  &14.7  & F5V     &0.136  &2.08  &n (E14) &n (L09)\\
201091  &61 Cyg A       & 21:06:53.95  &+38:44:58.0  &3.5   & K5V     &0.106  &4.43  &n (A13) &n (G04)\\  
201092  &61 Cyg B       & 21:06:55.26  &+38:44:31.4  &3.5   & K7V     &0.085  &3.28  &n (A13) &n (G04) \\ 
215648  &ksi Peg A      & 22:46:41.58  &+12:10:22.4  &16.3  & F7V     &0.132  &2.22  &n (E14) &n (G13)\\
219134  &               & 23:13:16.98  &+57:10:06.1  &6.5   & K3V     &0.080  &1.86  &        &n (T08)\\
222368  &iot  Psc       & 23:39:57.04  &+05:37:34.6  &13.7  & F7V     &0.137  &2.40  &n (MG11)&n (B06)\\ 
\enddata
\tablecomments{Excess References: A13=\citet{Absil:2013}; Be06=\citet{Beichman:2006};B06=\citet{Bryden:2006};
  B09=\citet{Bryden:2009}; E13=\citet{Eiroa:2013}; E14=\citet{Ertel:2014}; G13=\citet{Gaspar:2013}; G03=\citet{Greaves:2003}; G04=\citet{Greaves:2004}; K13=\citet{Kennedy:2013}; K10=\citet{Koerner:2010};L02=\citet{Laureijs:2002};   L09=\citet{Lawler:2009}; MG11=\citet{Millan-Gabet:2011};T08=\citet{Trilling:2008}}

\end{deluxetable*}

\begin{deluxetable*}{llrrccccll}
\tabletypesize{\scriptsize}
\tablecaption{Stars in the Sensitivity-Driven Sample\label{tab:bright}} 
\tablewidth{0pt}
\tablehead{
\colhead{HD}&\colhead{Name}&\colhead{RA} &\colhead{Dec}&\colhead{Distance}
                                                   &\colhead{Spectral} &\colhead{EEID}
                                                          &\colhead{F$_\nu$ (N-band)}  &\colhead{hot/warm}&\colhead{cold}\\
            &              &              &    &\colhead{(pc)}&\colhead{Type} &  &\colhead{(Jy)}&\colhead{excess}  &\colhead{excess}
}
\startdata
HD 33111   & bet Eri    & 05:07:51.0  & -05:05:11.2  & 27.4  & A3IV   &0.248 &3.72   &n (E14) &y (G13)\\
HD 38678   & zet Lep    & 05:46:57.3  & -14:49:19.0  & 21.6  & A2IV-V &0.176 &2.06   &y (FA98), n (A13) &y (FA98)\\
HD 40136   & eta Lep    & 05:56:24.3  & -14:10:03.7  & 14.9  & F2V    &0.161 &2.36   &        &y (L09)\\
HD 81937   & h UMa      & 09 31 31.7  & +63:03:42.8  & 23.8  & F0IV   &0.168 &2.55   &        &n (B06)\\
HD 95418   & beta UMa   & 11:01:50.5  & +56:22:56.7  & 24.4  & A1IV   &0.316 &4.20   &y (FA98), n (A13) &y (S06)\\
HD 97603   & del Leo    & 11:14:06.5  & +20:31:25.4  & 17.9  & A5IV   &0.278 &3.90   &n (A13) &n (G13)\\
HD 102647  & bet Leo    & 11:49:03.6  & +14:34:19.4  & 11.0  & A3V    &0.336 &6.85   &y (A13) &y (S06)\\
HD 103287  & gam Uma    & 11:53:49.8  & +53:41:41.1  & 25.5  & A1IV   &0.308 &3.69   &        &n (S06)\\
HD 105452  & alf Crv    & 12:08:24.8  & -24:43:44.0  & 14.9  & F1V    &0.139 &1.97   &        &n (G13)\\ 
HD 106591  & del UMa    & 12:15:25.6  & +57:01:57.4  & 24.7  & A2V    &0.199 &2.00   &n (A13) &n (G13)\\
HD 108767  & del Crv    & 12:29:51.8  & -16:30:55.6  & 26.6  & A0IV   &0.251 &2.25   &y (E14) &n (S06)\\
HD 109085  & eta Crv    & 12:32:04.2  & -16:11:45.6  & 18.3  & F2V    &0.125 &1.76   &n (A13) &y (W05)\\
HD 128167  & sig Boo    & 14:34:40.8  & +29:44:42.4  & 15.8  & F2V    &0.117 &1.39   &        &y (L02)\\
HD 129502  & 107 Vir    & 14:43:03.6  & -05:39:29.5  & 18.3  & F2V    &0.151 &2.60   &n (E14) &n (G13)\\
HD 164259  & zet Ser    & 18:00:29.0  & -03:41:25.0  & 23.6  & F2IV   &0.106 &1.14   &n (E14) &n (L09)\\                  
HD 172167  & Vega       & 18:36:56.3  & +38:47:01.3  &  7.7  & A0V    &0.916 &38.55  &y (A13) &y (G86)\\
HD 187642  & Altair     & 19:50:47.0  & +08:52:06.0  &  5.1  & A7V    &0.570 &21.63  &y (A13) &n (R05)\\
HD 203280  & Alderamin  & 21:18:34.8  & +62:35:08.1  & 15.0  & A8V    &0.294 &7.04   &y (A13) &n (C05)\\
HD 210418  & tet Peg    & 22:10:12.0  & +06:11:52.3  & 28.3  & A1V    &0.179 &1.61   &n (E14) &n (S06)\\
HD 216956  & Fomalhaut  & 22:57:39.0  & -29:37:20.0  & 7.7   & A4V    &0.504 &15.41  &y (L13) &y (G86)\\
\enddata
\tablecomments{Excess References: 
A13=\citet{Absil:2013}; 
B06=\citet{Bryden:2006}; 
C05=\citet{Chen:2005}; 
E14=\citet{Ertel:2014};
FA98=\citet{Fajardo:1998}; 
G13=\citet{Gaspar:2013}; 
G86=\citet{Gillett:1986};
L02=\citet{Laureijs:2002};
L09=\citet{Lawler:2009}; 
L13=\citet{Lebreton:2013}; 
R05=\citet{Rieke:2005}; 
S06=\citet{Su:2006}; 
W05=\citet{Wyatt:2005}}
\end{deluxetable*}

\begin{deluxetable*}{llll} 
\tabletypesize{\small}
\tablecaption{Binary Stars Excluded from the Sample\label{tab:binary}} 
\tablewidth{0pt}
\tablehead{
\colhead{HD}& \colhead{Name}    &\colhead{SpTyp} &\colhead{Binarity Notes}
}
\startdata
\hline
HD 432     &bet Cas   &F2IV   &WDS says SB with P=27d\\
HD 4614    &eta Cas   &G3V    &VB 12'', 70AU, P=480yr\\
HD 6582    &mu Cas    &G5V    &VB 1'', 7.5 AU + SB\\
HD 8538    &del Cas   &A5III-IV &SB, perhaps eclipsing\\
HD 11443   &alf Tri   &F6IV   &SB, P=1.8d\\
HD 11636   &bet Ari   &A5V    &SB9, P=107d\\
HD 13161   &          &A5IV   &SB, P=31d, resolved by MarkIII\\
HD 13974   &          &G0V    &SB, P=10d\\
HD 16970   &          &A2V    &VB, 2.3''\\
HD 20010   &          &F6V    &VB 4.4'', 62 AU\\
HD 20630   &kap01 Cet &G5V    &WDS says SB, but not variable in \citet{Nidever:2002}\\
HD 39587   &chi1 Ori  &G0V    &VB 0.7'', 6 AU, P=14yr\\
HD 40183   &          &A1IV-V &SB, P=3.7d, resolved by MarkIII\\
HD 47105   &          &A1.5IV &SB/speckle VB \\
HD 48915   &Sirius    &A0V    &VB, 7.5''\\
HD 56986   &del Gem   &F2V    &SB, P=6.1 yr\\
HD 60179   &Castor A  &A1.5IV &SB, P=9.2d plus VB, P=467 yr\\
HD 61421   &Procyon   &F5IV   &VB 5'', 18AU, P=40 yr and SB\\
HD 76644   &	      &A7V    &SB, P=11yr\\
HD 76943   &10 UMa    &F3V    &SB/VB 0.6'', P=21.8 yr\\
HD 82328   &tet UMa   &F7V    &WDS says SB (SBC7), but not in SB9, also VB, 5''\\
HD 82885   &SV Lmi    &G8III  &VB 3.8'', 43 AU, P=200 yr\\
HD 95735   &          &M2V    &EB\\
HD 98231   &GJ 423A   &G0V    &Complicated multiple system\\
HD 104304  &          &G8IV   &VB, 1''\\
HD 109358  &bet Cvn   &G0V    &SB, 0.1'', 1.3 AU\\
HD 110379  &GJ 482A   &F0V    &VB, 3.75'', 44AU, P=171yr\\
HD 112413  &alf02 CVn &A0II-III &SB\\
HD 114378J &alph Com  &F5V    &VB 0.7'', 12 AU, P=26yr\\
HD 114710  &bet Com   &G0V    &possible SB\\
HD 116656  &Mizar A   &A1.5V  &SB, P=20.5d\\
HD 118098  &zet Vir   &A3V    &VB, companion is M4-M7\\
HD 121370  &eta Boo   &G8V    &SB9, P=494d\\
HD 130841  &alf Lib A &A4IV-V &possible SB\\
HD 131156  &37 Boo    &G8V    &VB 4.9'', 33AU, P=151yr\\
HD 131511  &GJ 567    &K2V    &possible SB P=125d\\
HD 133640  &GJ 575    &G0V    &VB, 3.8'', 48 AU (now at 1.5'')\\
HD 139006  &alf CrB   &A1IV   &SB, P=17d\\
HD 140538  &          &G2.5V  &VB, 4.4'', 65 AU\\
HD 144284  &tet Dra   &F8 IV-V   &SB, P=3.1d\\
HD 155125  &          &A2IV-V &VB 0.86''\\
HD 155886  &GJ 663A   &K2V    &VB 5'', 28 AU\\
HD 155885  &36 Oph    &K2V    &VB 15'', 87AU, P=569yr + possible SB\\
HD 156164  &del Her   &A1IV   &SB 0.1''\\
HD 156897  &40 Oph    &F2V    &VB, approx 4'' (CCDM)\\
HD 159561  &alf Oph   &A5II   &VB 0.8''\\
HD 160269  &          &G0V    &VB 1.5'', 21 AU\\
HD 160346  &GJ 688    &K3V    &SB P=84d\\
HD 161797  &mu Her    &G5IV   &VB 1.4'', 12AU, P=65yr\\
HD 165341  &70 Oph A  &K0V    &VB 4.6'', 23AU, P=88yr\\
HD 165908  &b Her     &F7V    &VB 1.1'', 17 AU, P=26 yr\\
HD 170153  &          &F7V    &SB 0.1'', 1AU, P=0.8yr\\
CCDM 19026-2953 A &   &A2.5V  &VB 0.53''\\
HD 177724  &          &A0IV-V &SB + VB, 5''\\
HD 182640  &del Aql   &F1IV-V &SB, resolved at 0.1''\\
HD 185395  & tet Cyg  &F4V    &VB/SB, approx 2.5''\\
HD 207098  & del Cap  &F2III  &EB\\
HD 210027  &iot Peg   &F5V    &SB1, P=10d\\
HD 224930  &85 Peg    &G5V    &VB 0.8'', 10AU, P=26 yr\\
\enddata
\end{deluxetable*}

\bibliographystyle{apj} 
\bibliography{lbti_paper} 
 
\end{document}